\documentclass[preprint,12pt]{elsarticle}
\usepackage{amssymb}
\usepackage{amsmath}
\usepackage{algorithm}
\usepackage{algpseudocode}
\usepackage{float} %
\usepackage{setspace}
\usepackage{color}
\usepackage{siunitx}

\begin{document}

\begin{frontmatter}

\title{Accelerating Gaussian beam tracing method with dynamic parallelism on graphics processing units} 

\author[1,3]{Zhang Sheng}
\author[1,4]{Lishu Duan\corref{cor1}}
\author[1,2]{Hanbo Jiang\corref{cor2}}

\affiliation[1]{organization={Eastern Institute for Advanced Study, Eastern Institute of Technology},
                city={Ningbo},
                country={China}}

\affiliation[2]{organization={Ningbo Institute of Digital Twin, Eastern Institute of Technology},
                city={Ningbo},
                country={China}}

\affiliation[3]{organization={School of Mathematics, Hefei University of Technology},
                city={Hefei},
                country={China}}

\affiliation[4]{organization={School of Ocean and Civil Engineering, Shanghai Jiao Tong University},
                city={Shanghai},
                country={China}}
                
\cortext[cor1]{Corresponding author. Email: duanlishu@sjtu.edu.cn}

\cortext[cor2]{Corresponding author. Email: hjiang@eitech.edu.cn}

\begin{abstract}

This study presents a reconstruction of the Gaussian Beam Tracing solution using CUDA, with a particular focus on the utilisation of GPU acceleration as a means of overcoming the performance limitations of traditional CPU algorithms in complex acoustic simulations. The algorithm is implemented and optimised on the NVIDIA RTX A6000 GPU, resulting in a notable enhancement in the performance of the Gaussian Beam Summation (GBS) process. In particular, the GPU-accelerated GBS algorithm demonstrated a significant enhancement in performance, reaching up to 790 times faster in city enviroment and 188 times faster in open plane enviroment compared to the original CPU-based program. To address the challenges of acceleration, the study introduce innovative solutions for handling irregular loops and GPU memory limitations, ensuring the efficient processing of large quantities of rays beyond the GPU's single-process capacity. Furthermore, this work established performance evaluation strategies crucial for analysing and reconstructing similar algorithms. Additionally, the study explored future directions for further accelerating the algorithm, laying the groundwork for ongoing improvements.

\end{abstract}


\end{frontmatter}

\section{Introduction}
The study of sound propagation is crucial for a wide range of applications, including the design of concert halls~\cite{allred1958applications,barron2009auditorium}, 
the enhancement of virtual reality experiences~\cite{vorlander2014virtual}, improvements in audio prediction for gaming~\cite{raghuvanshi2009efficient,tan2023virtual}, and the assessment of environmental noise from drones and unmanned aerial vehicles~\cite{bian2022efficient,tan2024enhancing,tan2024low}.
One of the key challenges in acoustics simulation is modeling the interaction of sound waves with complex environments, which involves reflections, diffractions, and scattering. 

Over the past few decades, various simulation approaches have been developed to address these challenges, broadly classified into wave-based and geometrical acoustics~\cite{eaton1996application,hart2011prediction,deines2006comparative}. Wave-based methods,directly solve the wave equation to capture phenomena like interference and diffraction,while the traditional numerical method such as finite element method~\cite{eaton1996application} and finite different method~\cite{jiang2019reduced} are adopted. Despite these methods provide good accuracy, their substantial computational demands make them inappropriate for large-scale applications with complex geometries.
Conversely, geometrical acoustics simplifies sound as rays or beams and models reflections in a computationally efficient manner, making it particularly useful for large environments~\cite{laine2009accelerated}.  
In this regard, the image source method and ray tracing method are widely used in various acoustic applications. The image source method~\cite{allen1979image} models each reflection from a surface as if it originates from an image source symmetrically positioned on the opposite side of that surface. 
This method is highly effective for simpler enviroment such as predicting room impulse responses, analyzing reverberation, and optimizing sound design in spaces like auditoriums and studios.
However, as the order of reflections increases, particularly in environments with dense occlusions, the number of image sources grows exponentially, leading to a significant increase in computational complexity~\cite{laine2009accelerated}. 

On the other hand, the ray tracing method represents sound as a collection of rays that emanate from a source~\cite{lehnert1993systematic}. These rays travel through the environment and interact with surfaces encountered.
Various versions of the ray tracing algorithm have been implemented ever since the pioneering work of Krokstadt et al.~\cite{krokstad1968calculating}. 
However, the standard ray tracing method suffers perfect shadows and caustics~\cite{porter1987gaussian}. To address these limitations, the Gaussian beam tracing(GBT) method was introduced as an extension to traditional ray tracing method~\cite{porter1987gaussian}. 
This approach associates each ray with a beam that has a Gaussian intensity distribution then constructs the sound at any given point by summing the contributions of each beam~\cite{porter1994finite}. 
Unlike traditional rays, Gaussian beams inherently spread sound energy over a continuous area, filling in gaps where traditional rays might fail. The improvements are twofold:
Physically, Gaussian beams allow for smooth transitions in sound intensity across space, preventing infinite energy accumulation in caustic regions and reducing unwanted spikes or dips in sound pressure~\cite{porter1987gaussian}. 
Numerically, Gaussian beams enhance sound field coverage while requiring fewer computational resources. 
This proves particularly advantageous in complex environments with numerous reflections and refractions~\cite{bian2022efficient}, as it reduces artifacts that would otherwise demand significant post-processing.

Nevertheless, GBT method still faces challenges, particularly in large environments with complex geometries, where the number of beams and interactions increases significantly, leading to higher computational demands. To address these limitations, parallel computation can be employed. It distributes computational tasks across multiple processing units, thereby reducing the time required to trace beams, compute interactions, and reconstruct the sound field. Among parallel architectures, graphics processing units (GPUs) have proven particularly effective due to their ability to handle many tasks simultaneously.
GPUs are equipped with thousands of cores that can process numerous parallel tasks concurrently, making them well-suited for data-intensive applications and independent tasks~\cite{he2009relational}. This architecture provides significant speedups for highly parallel tasks~\cite{navarro2014survey}, such as those encountered in geometric acoustics.
Spjut et al. investigated a multi-threaded beam tracing algorithm on multicore platforms, achieving significant speedups with an increased number of threads~\cite{spjut2009trax}. Cowan and Kapralos discussed GPU ray tracing techniques and demonstrated performance improvements for real-time acoustic prediction~\cite{cowan2010gpu}. Gkanos et al. implemented the image source method on multiple GPUs and suggested that incorporating beam tracing could further enhance efficiency~\cite{gkanos2021comparison}. Tan et al. proposed a GPU-based tree-accelerated beam-tracing method, achieving a speedup of 66 times compared to conventional techniques~\cite{tan2015full}. 
Additionally, Greef et al. employed a ray tracing algorithm for radiotherapy dose calculations on a GPU, achieving a 6 times speedup for the evaluated cases~\cite{de2009accelerated}.

While significant advancements have been made in accelerating Ray Tracing(RT) method, which is only a part of GBT, the reconstruction of the sound pressure field remains underexplored. This crucial process involves searching for all beams contributing to the sound pressure at a specific field point. Existing approaches do not fully leverage parallel computation capabilities, leading to inefficiencies. This study proposes the use of dynamic parallelism on GPUs, which allows for on-the-fly parallel task generation~\cite{guide2013cuda}. By enabling kernels to launch other kernels, this method allows threads to manage their parallelism dynamically without the need for explicit synchronization through the host. This approach not only reduces the overhead of kernel launches but also improves overall resource utilization, which is particularly beneficial for collecting contributions from all Gaussian beams where the computational load varies from point to point~\cite{cope2010performance,czarnul2003programming}.

The remainder of this paper is organized as follows: Section 2 introduces the fundamental principles of Gaussian Beam Tracing (GBT), including its mathematical framework and the computational challenges encountered in large-scale acoustic simulations. Section 3 details the development of the GPU-accelerated algorithm, emphasizing the integration of flat and dynamic parallelism to address bottlenecks in computational performance. The implementation strategies, including memory management and workload balancing, are also discussed. Section 4 presents the validation of the proposed method against analytical solutions, followed by a comprehensive performance evaluation in different acoustic environments. Section 5 explores numerical experiments conducted in both free-field and city environments, highlighting the significant performance gains achieved with the optimized GPU algorithm and identifying the remaining limitations. Finally, Section 6 concludes the paper by summarizing the key findings and proposing future directions for further research and optimization.

\section{Gaussian beam tracing \label{s:Method}}
\begin{figure}
\centering
\includegraphics[width=0.5\linewidth]{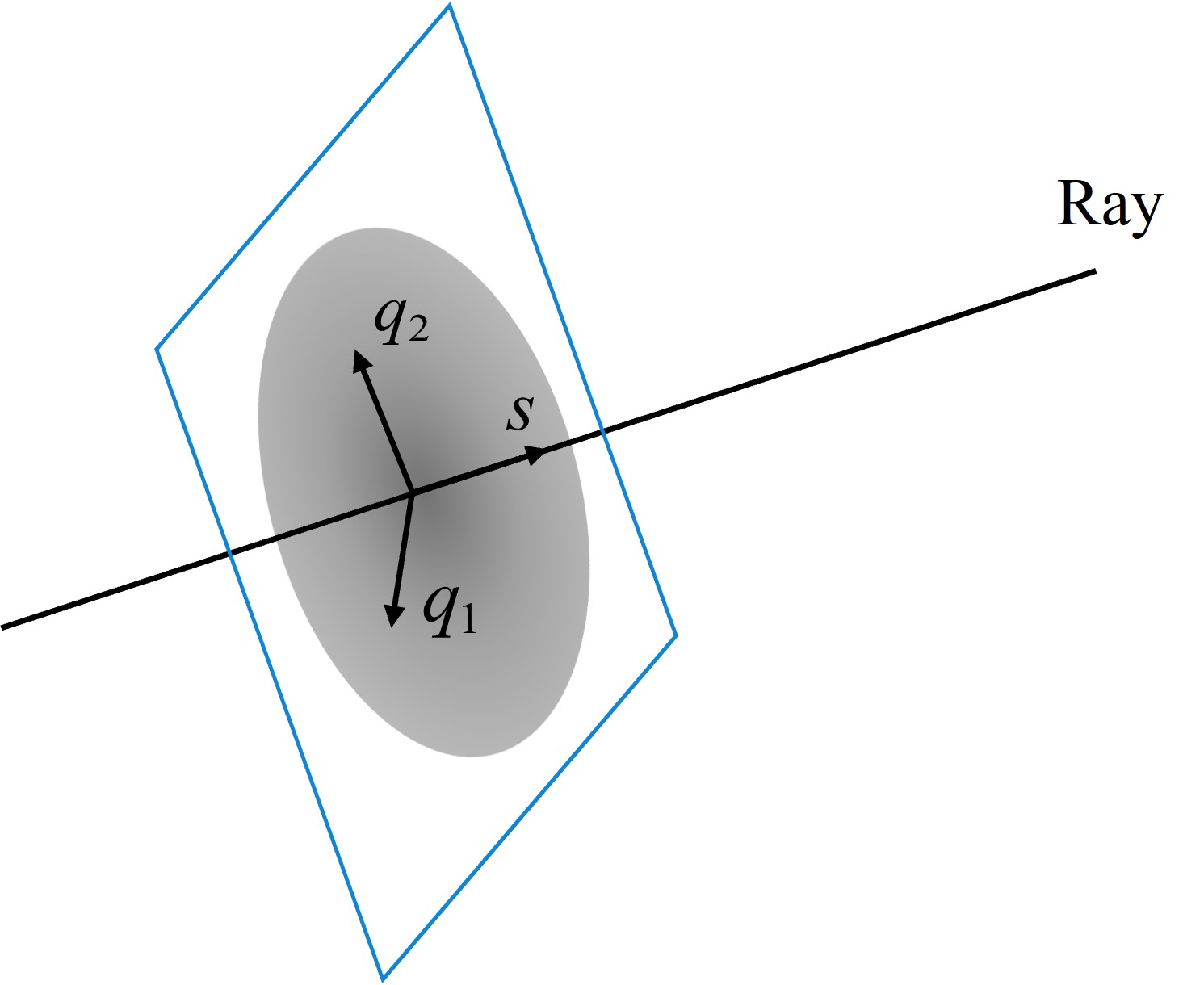} 
\caption{\label{fig:beamModel}
Schematic of a Gaussian beam and the associated coordinates.
}
\end{figure}
Figure~\ref{fig:beamModel} illustrates the Gaussian beam model, where the ray represents the path of sound propagation and is typically computed using a ray tracing algorithm. The Gaussian-shaped energy distribution around the ray is given by the following expression:
\begin{align}
 p(s,q_1,q_2,t) = &
 \, \phi\left(\frac{C(s)}{\text{det}[Q(s)]}\right)^{{1}/{2}} \notag \\
 & \times \exp\left[
 -\text{i}\omega\left(t-\int_s^{s_0}\frac{\text{d}s}{C(s)}\right)
 + \frac{\text{i}\omega}{2}\left(\mathrm q^T P Q^{-1}\mathrm q\right)
 \right],
\end{align}
where $\phi$ is a real constant, and $(s, q_1, q_2)$ represent the ray-centered coordinates.
Here, $\text{det}[\cdot]$ represents the determinant operator, and $P$ and $Q$ are matrices that satisfy the following relations:
\begin{equation}
\frac{\partial\mathrm{Q}}{\partial s}=c\mathrm{P},\quad\frac{\partial\mathrm{P}}{\partial s}=0,
\end{equation}
where $c$ is the sound speed.
The method for calculating ray paths has been extensively studied and will therefore not be repeated here. Interested readers are encouraged to refer to the relevant literature for more detailed implementations~\cite{Speer1992An}. Subsequently, the contributions of all nearby Gaussian beams along each ray are gathered to estimate the sound pressure $p$ at any observation point $R$
\begin{equation}
\label{e:GBSEquation}
p(R,\omega) = 
\int\int \Phi(\gamma_1, \gamma_2) P(R_\gamma, \omega) \exp[\text{i}\omega T(R, R_\gamma)] \text{d}\gamma_1 \text{d}\gamma_2,
\end{equation}
where  $\Phi(\gamma_1, \gamma_2)$ is the weighting function, $P(R_\gamma, \omega)$ represents the complex amplitude along the ray $R_\gamma$, and $T(R, R_\gamma)$ is the propagation time from the point $ R_\gamma$ to $R$.
In this formula, another important coordinates called ray coordinates $(s, \gamma_1, \gamma_2)$ 
is presented, which are connected to the whole ray field, and $\gamma_1$ and $\gamma_2$ are the parameters of the ray at the source.
For more details on the GBT method, please refer to refer to the literature~\cite{bian2022efficient}. 
It is worth noting that numerically evaluating the above integration is time-consuming, but this process can be significantly accelerated using GPU calculations.

\section{Algorithm and implementation\label{s:Implementation}}

\subsection{Overview}
\begin{figure}
\centering
\includegraphics
[width=1.0\linewidth]
{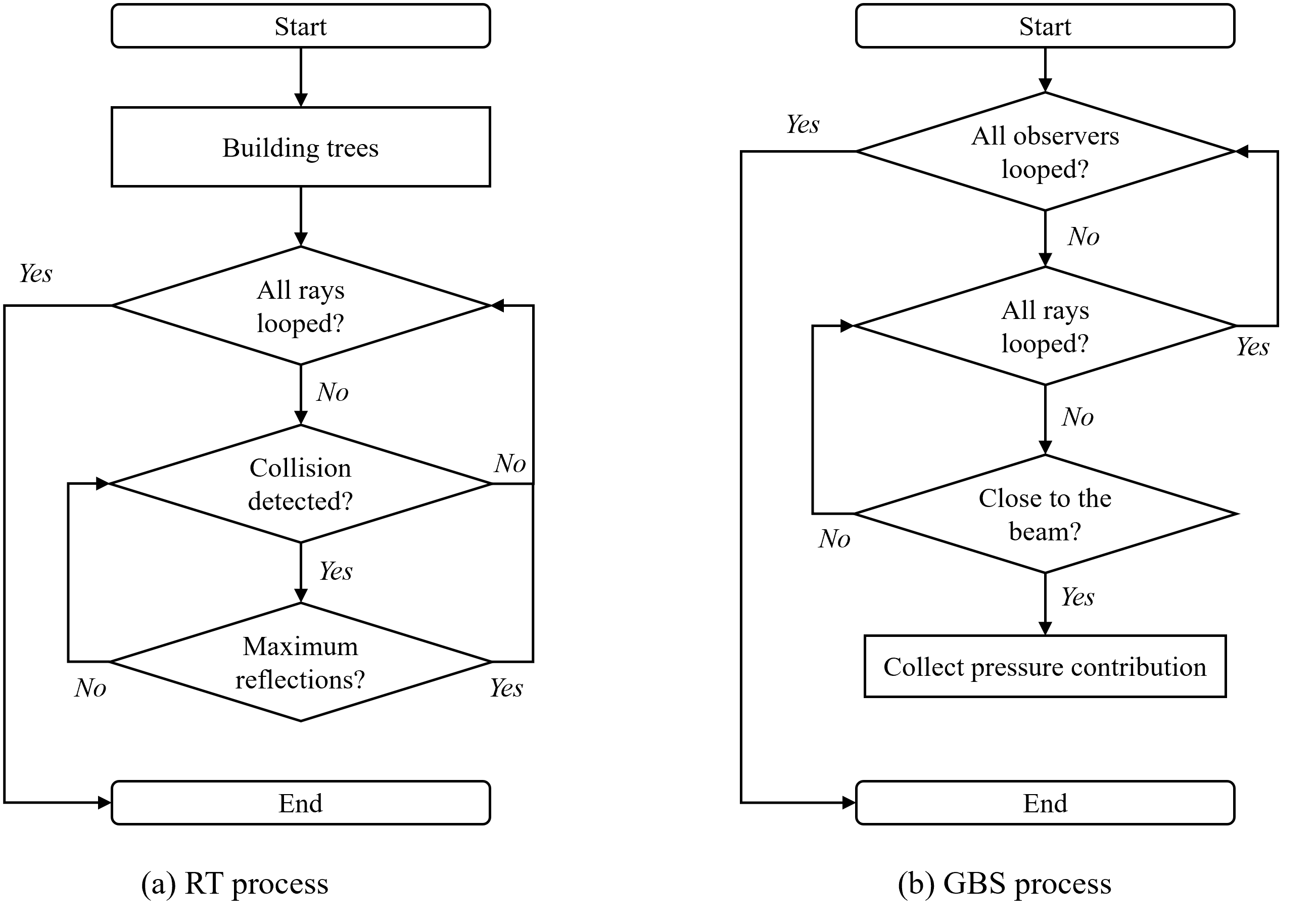}
\caption{
The programming flowcharts of the RT and  GBS processes. The proposed dynamic parallelism targets to accelerate looping all rays in each process.
\label{fig:RT_GBS_flowchart}
}
\end{figure}
This section presents the numerical algorithm for Gaussian beam tracing and its parallel implementation. The overall workflow consists of two primary processes: ray tracing (RT) and Gaussian beam summation (GBS), as illustrated in Fig.~\ref{fig:RT_GBS_flowchart}.
In the RT process, sound propagation paths are determined by tracing rays and modeling their interactions with obstacles. The complexity of this stage is driven by the number of rays and boundary elements, where the latter are typically triangular facets representing reflective surfaces.
Subsequently, the GBS process aggregates contributions from all Gaussian beams to calculate sound pressure at observation points. Its computational demand is primarily influenced by the number of rays and observation points.

\subsection{CUDA architecture}
\begin{figure}
\centering
\includegraphics[width=1.0\linewidth]{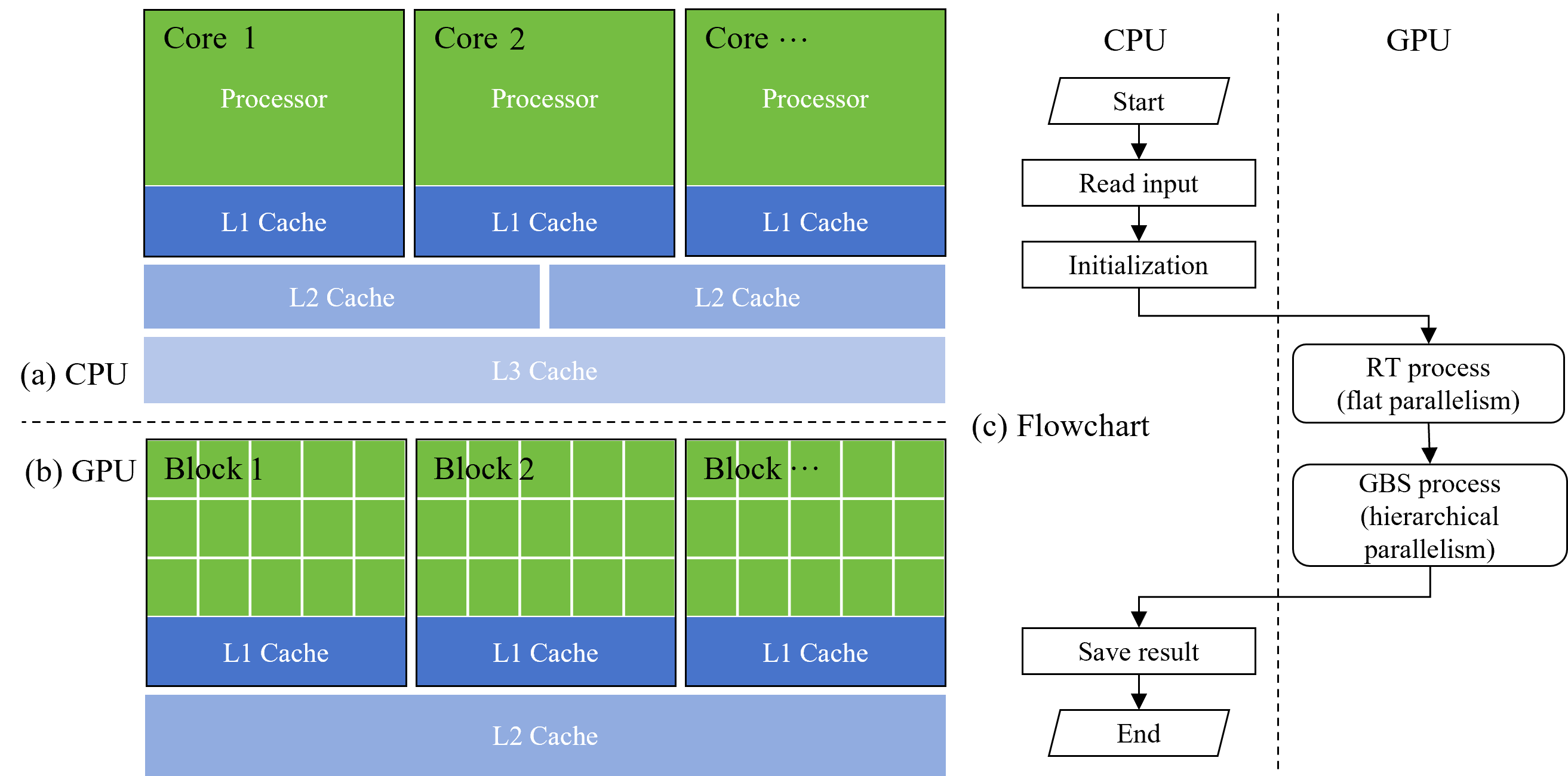}
\caption{(a) Typical CPU architecture; (b) Typical GPU architecture; Green indicates the processor while blue represents the memory; (c) Flowchart of the GPU acceleration implementation.
\label{fig_GPUStructure}
} 
\end{figure}
We now turn to the numerical implementations utilizing the compute unified device architecture (CUDA), a parallel computing platform and application programming interface developed by NVIDIA$^\circledR$. 
The CUDA programming model enables us to leverage computational resources from both CPU and GPU platforms.
Fig.~\ref{fig_GPUStructure}(a) illustrates the memory hierarchy in CPUs, consisting of three levels of cache memory: L1, L2, and L3. L1 performs the smallest and fastest, directly integrated into each core; L2 denotes larger and slower, dedicated to each core or shared among a few cores; and L3 represents the largest and slowest, shared across all cores, acting as a buffer between the cores and the main memory (random access memory, RAM).
In contrast, the GPU functions as a computation grid, consisting of hundreds of blocks, each containing thousands of processors, as shown in Fig.~\ref{fig_GPUStructure}(b). Threads within these blocks run concurrently, processing data in parallel using shared L1 cache memory. Each block can execute cooperatively via barrier synchronization. However, blocks typically do not share data directly with one another, except through global memory (L2) or other memory structures.

The proposed GBT implementation consists of both sequential and parallel components. As shown in Fig.~\ref{fig_GPUStructure}(c), tasks with low degrees of parallelism, such as file input/output operations, are assigned to the CPU to fully leverage its strengths and efficiency in sequential processing. Conversely, tasks like the RT and GBT processes, which involve high degrees of parallelism and heavy computational loads, are offloaded to the GPU which excels at handling numerous parallel operations simultaneously. Once these computations are completed, the results are transferred back to the CPU, where the remaining output and prediction processes continue in the same manner as the traditional CPU algorithm. 
This division of workflow ensures optimal utilization of each processor based on the nature of the tasks.

 \subsection{Flat parallelism \label{s:Flat_parallelism}}
\begin{figure}
\centering
\includegraphics
[width=0.8\linewidth]{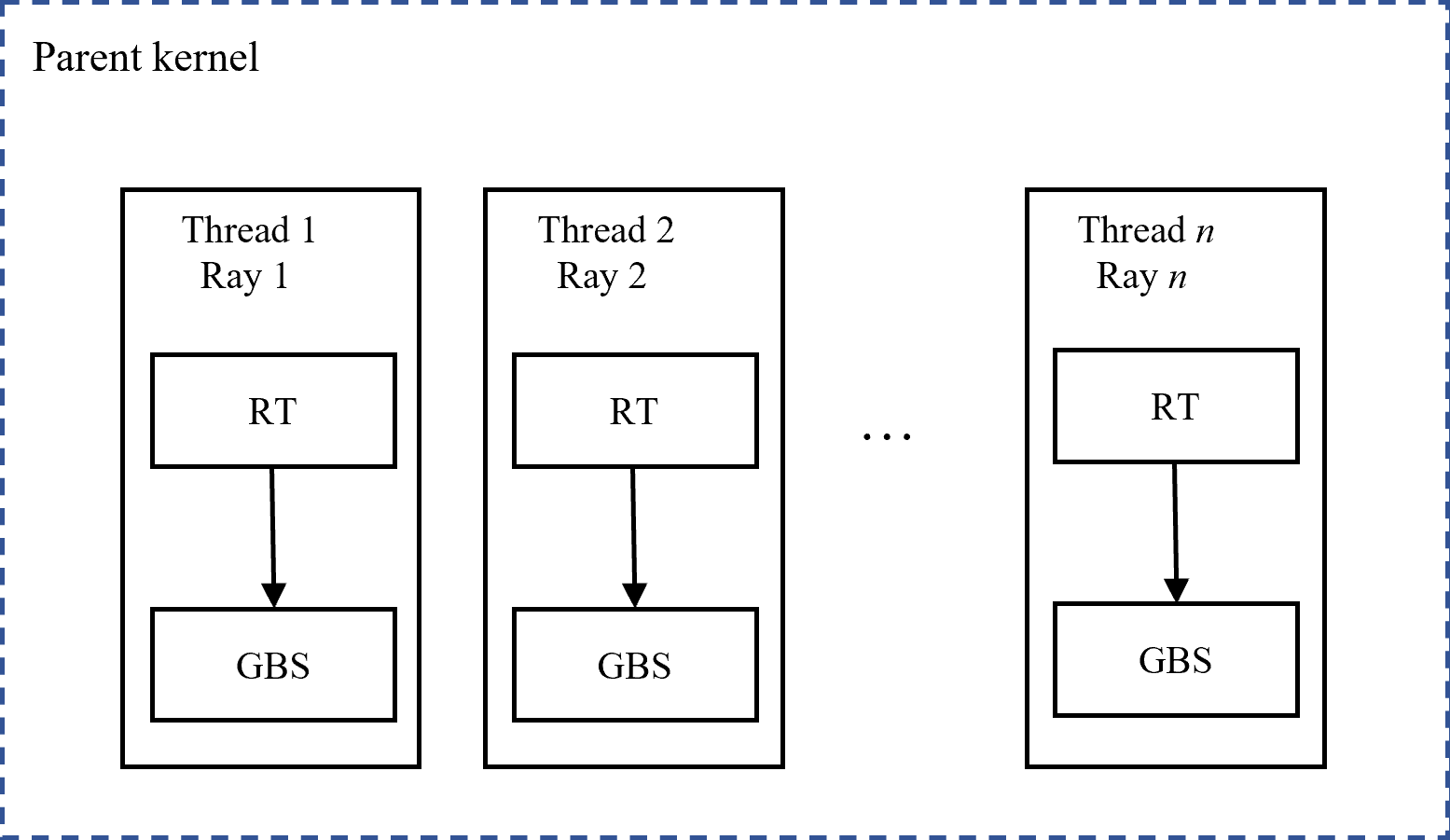}
\caption{
The multi-threaded programming model of flat parallelism on GPU. These threads run simultaneously.
\label{fig:Flat_parallelism}    
}
\end{figure}
Figure.~\ref{fig:Flat_parallelism} illustrates the flowchart of flat parallelism, detailing the entire algorithmic process of executing the RT and GBS steps.
The algorithm handles an array of rays, with the initial direction of each ray defined by specified combinations of elevation and azimuth angles. Each ray’s computation, encompassing both RT and GBS processes, is assigned to a dedicated thread. 
By evenly distributing these tasks across processors, flat parallelism fully exploits the computational capabilities of GPUs. After all rays are looped, the GPU computation finished and the whole workflow enters CPU phase.

\subsection{Dynamic parallelism \label{s:Dynamic_parallelism}}
\begin{figure}
\centering
\includegraphics
[width=0.7\linewidth]{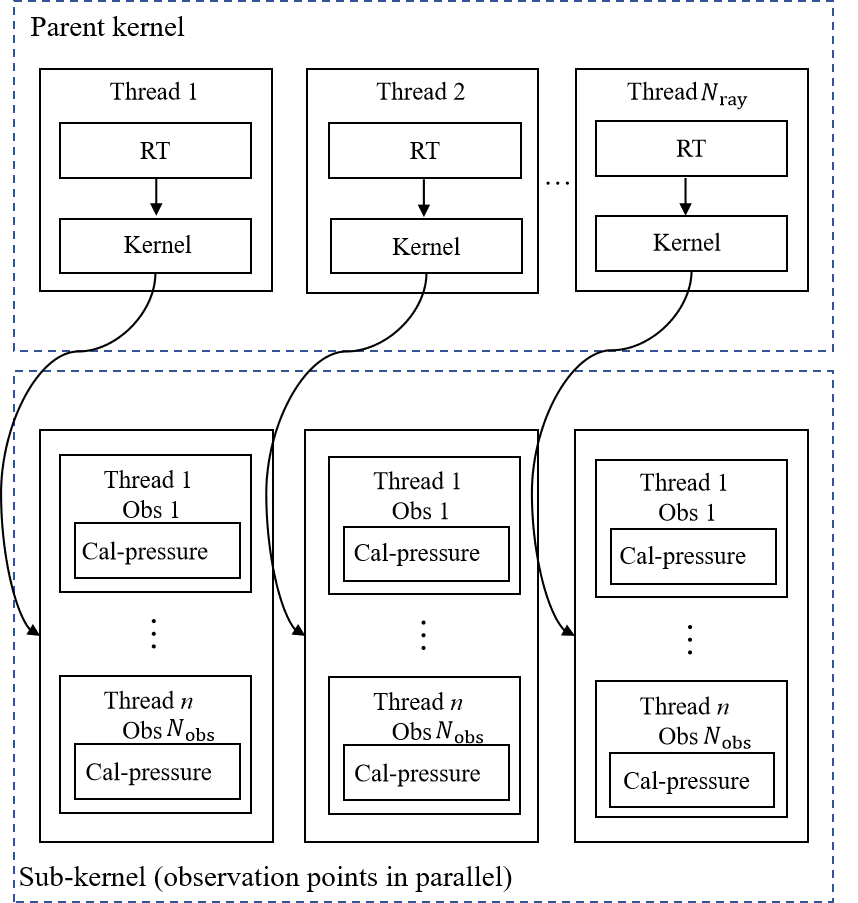}
\caption{
The multi-threaded programming model of dynamic parallelism on GPU.
\label{fig:Dynamic_parallelism} 
}
\end{figure}
In GBT process, each beam appears to be independent, making it well-suited for flat parallelism. However, the computational load varies across different observers and beams. This variation arises from the need to loop through all beams that contribute to the sound pressure in the GBS process, as described by Eq.~(\ref{e:GBSEquation}). The length of each beam can vary significantly, especially in complex environments where sound propagation is more intricate. Using flat parallelism resulted in uneven execution times: lightly loaded threads completed quickly, leaving heavy tasks to dominate the overall runtime. 

To address these inefficiencies, the present study adopts dynamic parallelism, a technology introduced by NVIDIA~\cite{guide2013cuda}, which allows kernels to launch additional kernels during execution, as illustrated in Fig.~\ref{fig:Dynamic_parallelism}. This hierarchical execution model enables fine-grained control over task allocation and workload balancing. Specifically, when a thread encounters a computationally intensive task, it can dynamically spawn child kernels to divide the workload further. These child kernels are executed independently, allowing idle GPU cores to be reallocated in real time to assist with heavy tasks. This mechanism ensures that computational resources are utilized efficiently, reducing idle time and mitigating bottlenecks.

\begin{figure}
\centering
\includegraphics
[width=1.0\linewidth]
{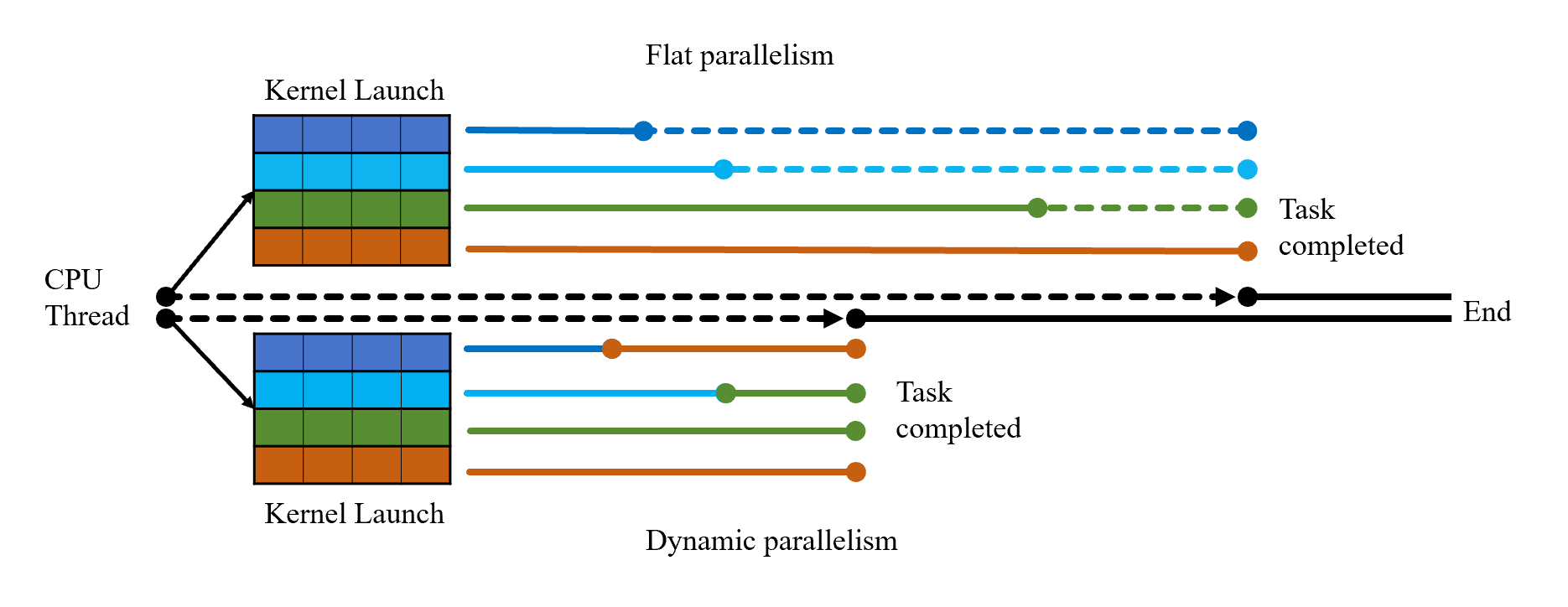}
\caption{Comparison between flat and dynamic parallelisms.
\label{fig:ParallelismComparison} 
}
\end{figure}

Figure.~\ref{fig:ParallelismComparison} compares flat and dynamic parallelism. In flat parallelism, tasks are statically distributed across GPU threads during kernel launches. This approach often leads to inefficiencies, as lightly loaded threads finish early and remain idle, while heavily loaded threads create bottlenecks. In contrast, dynamic parallelism redistributes the workload dynamically, enabling idle cores to assist heavily loaded threads. This adaptive rebalancing accelerates task execution significantly reduces overall computation time. The transition from flat to dynamic parallelism is particularly advantageous in environments with highly non-uniform task distributions, ensuring better load balancing and more efficient resource utilization.

\section{Results and discussion}
\subsection{Verification \label{s:verification}}
\begin{figure}
\centering
\includegraphics[width=1.0\linewidth]{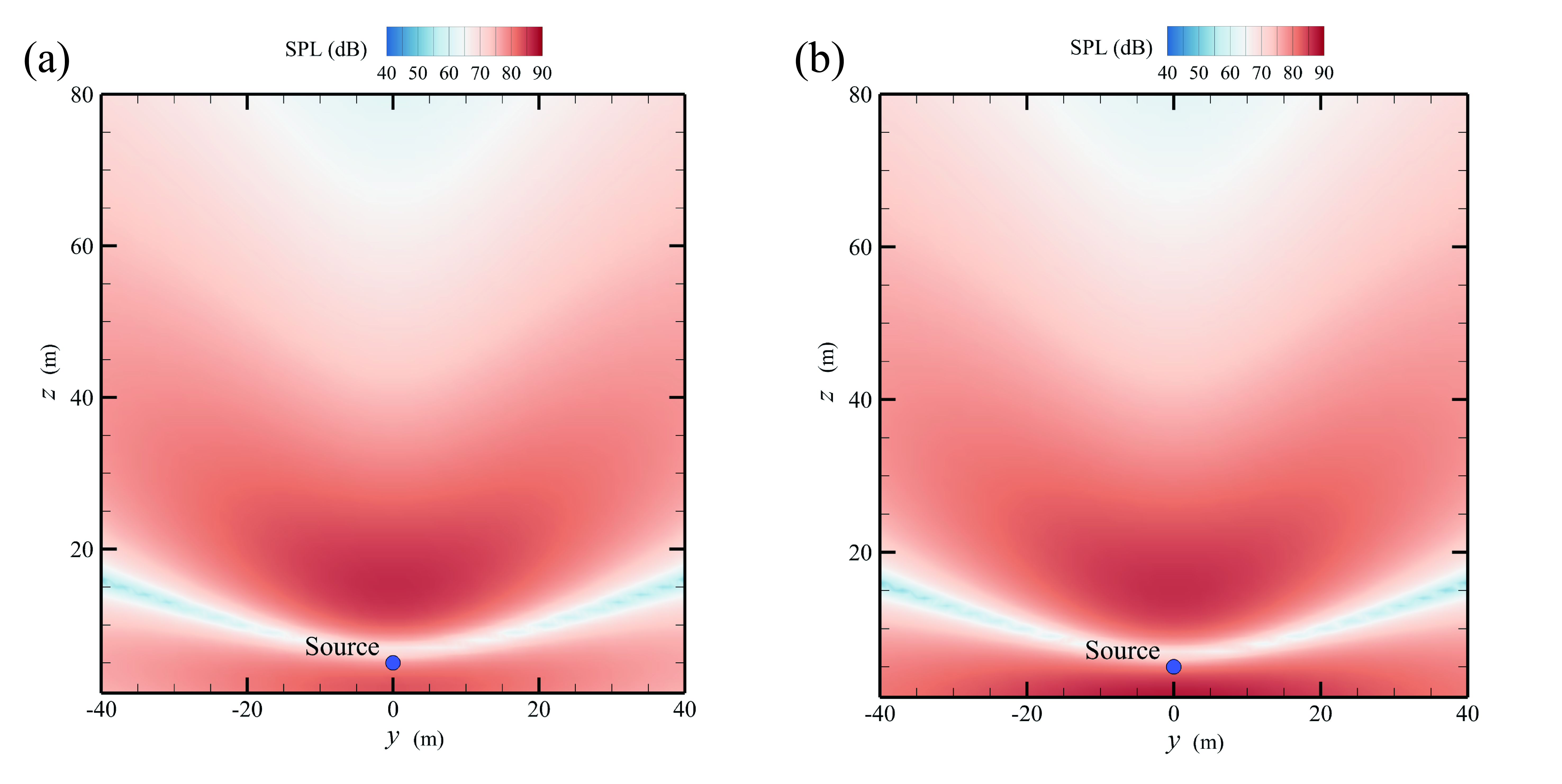}
\caption{
Comparison of whole SPL calculated by (a) GBT method and (b) analytical solution at $f=50 \text{Hz}$. 
\label{fig:verification_1-f50}
}
\end{figure}
\begin{figure}
\centering
\includegraphics[width=1.0\linewidth]{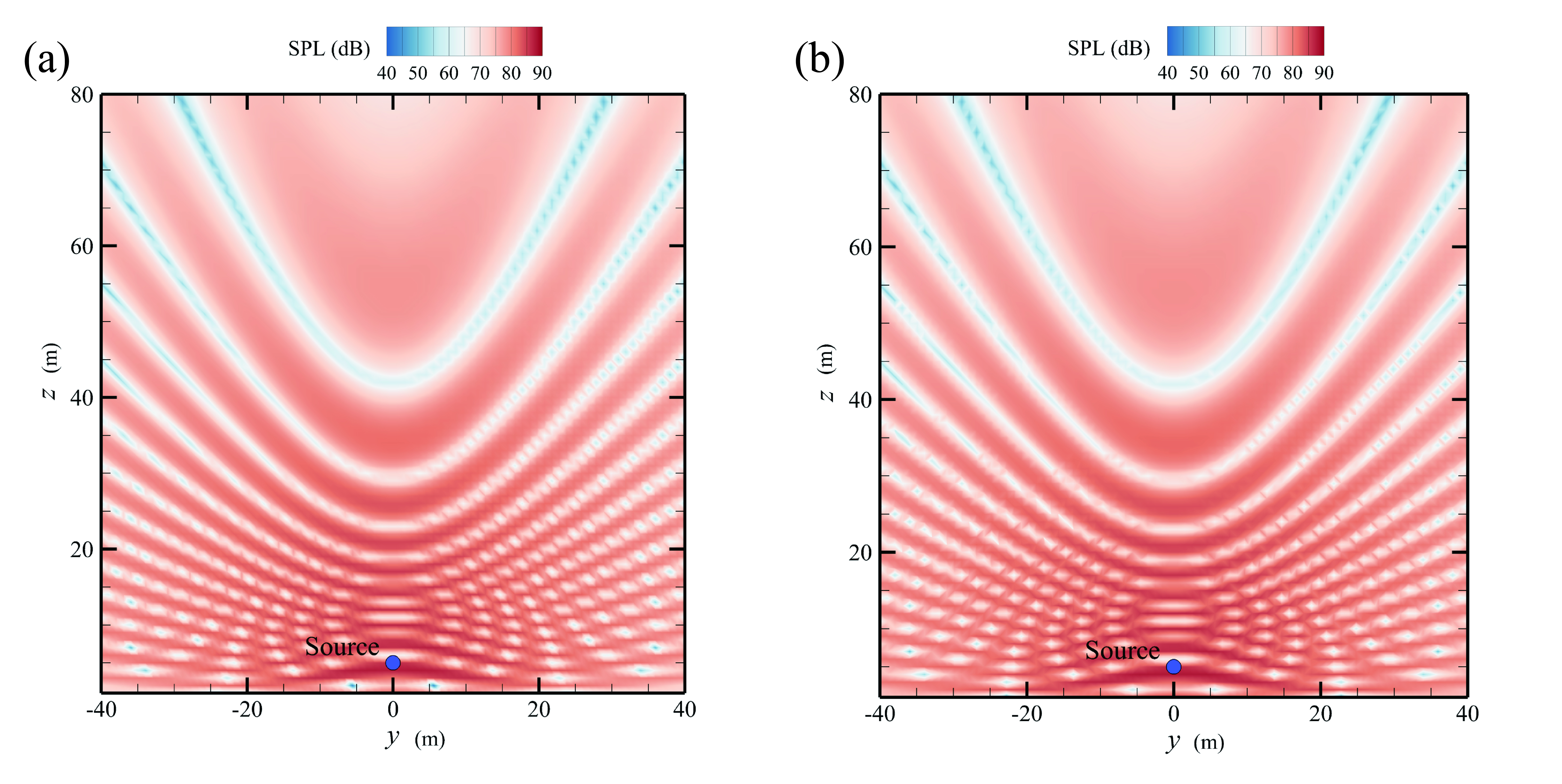}
\caption{
Comparison of whole SPL calculated by (a) GBT method and (b) analytical solution at $f=500 \text{Hz}$. 
\label{fig:verification_1-f500}
}
\end{figure}
\begin{figure}
\centering  
\includegraphics[width=1.0\linewidth]{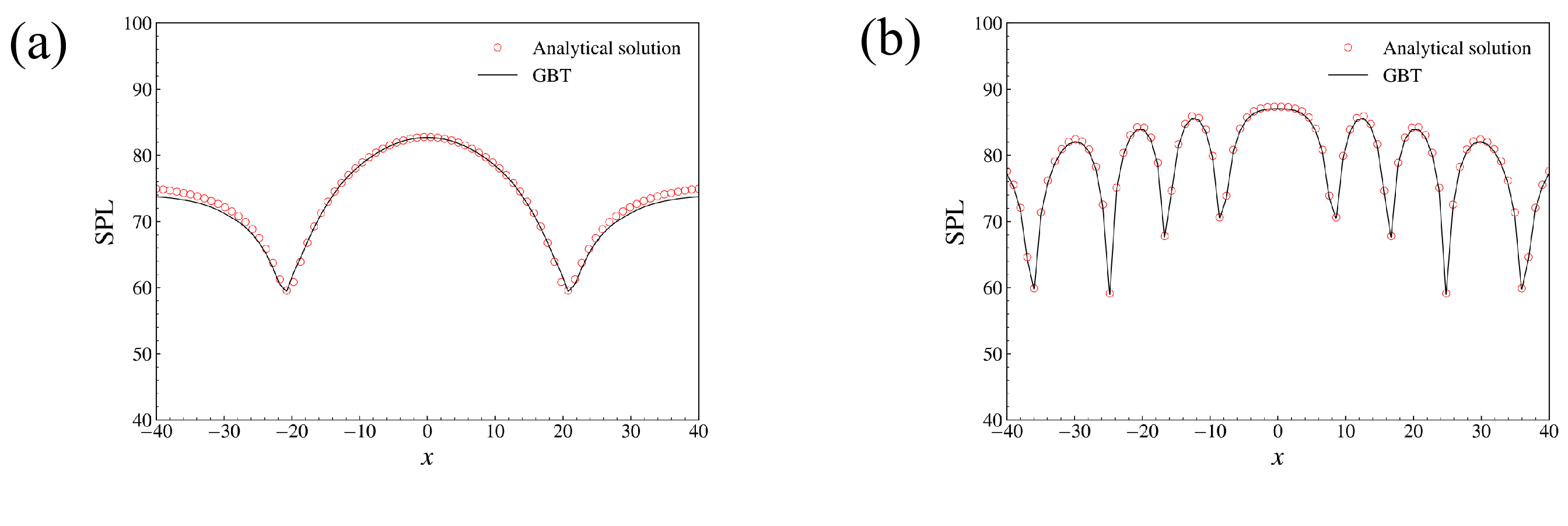}
\caption{
Comparison of SPL at $z=10$ using dynamic range with error bars for various frequencies: (a) 50\text{Hz}, (b) 500\text{Hz}.}
\label{fig:verification_2}
\end{figure}

To verify the correctness of the method described in this paper, the reflection of a monopole sound source above an infinitely long plate was computed. The results obtained by the GBT method are compared with the analytical solution. The sound source is positioned at $z=5$ with a frequency of 500Hz.

Figure.~\ref{fig:verification_1-f50} compares the sound field of a monopole sound source positioned above a rigid ground at $z=5$m with a frequency of 50Hz. The results demonstrate that the GBT solver reliably reproduces the sound field structure, as validated against the analytical solution.Fig.~\ref{fig:verification_1-f500} also illustrates the case for a sound source frequency of 500Hz, where the GBT solver's results continue to closely match the analytical solution. The comparison further confirms the solver's accuracy across different frequencies.Fig.~\ref{fig:verification_2} compares the sound pressure level distribution at $z=10$, showing an excellent match between the two results.

This validation case demonstrates the accuracy of the proposed method and its implementation even for high-frequency sound reflection problems, providing a strong foundation for subsequent calculations.

\subsection{Application to environmental noise}
\begin{figure}
    \centering
    \includegraphics[scale=0.23]{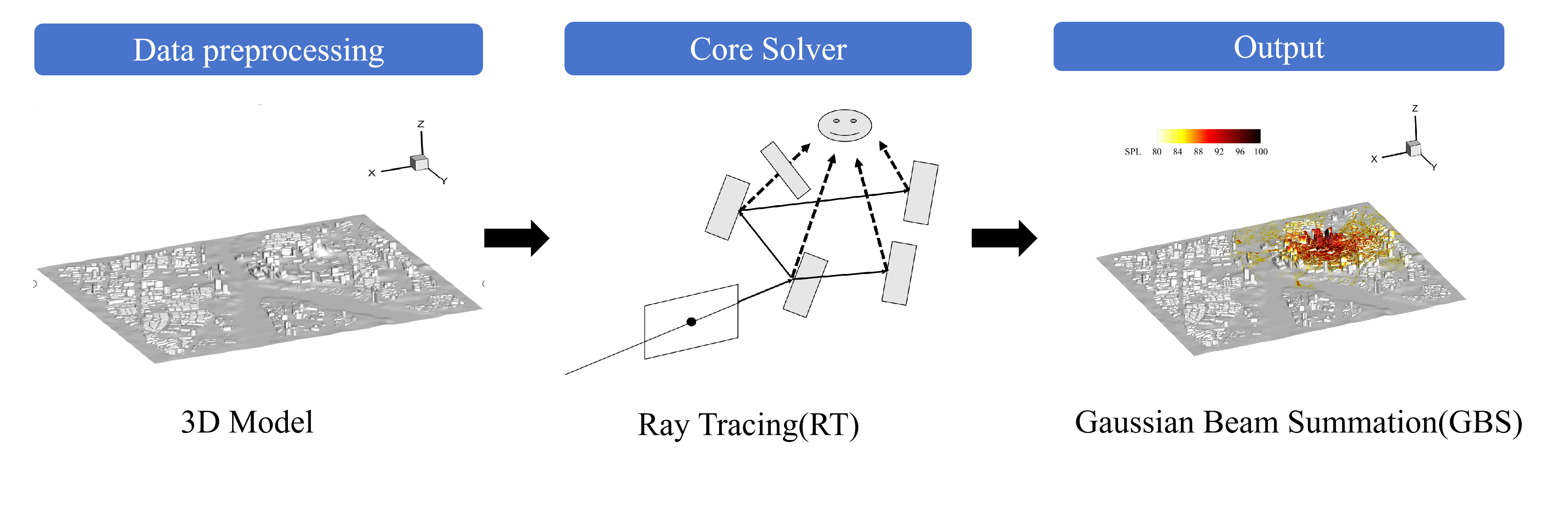}
    \caption{Single-threaded sound field prediction algorithm framework.}
    \label{fig:Single-threaded——diagram}
\end{figure}
In the GBT algorithm, key information such as the characteristics of the noise source, the geometric environment model, and the observer's position are taken as inputs, which then transforms the geometric objects in the inputs into structures optimized for accelerated processing. Subsequently, the RT module are responsible for executing two core tasks: calculating the propagation trajectories of each Gaussian beam and predicting the sound pressure contributions of these beams along their propagation paths. For each observation point, the final calculation of sound pressure is completed through the GBS module. This method involves integrating the contributions of all Gaussian beams near the observation point to construct the complete sound field. This entire computational process is detailed in Fig.~\ref{fig:Single-threaded——diagram}.

\subsection{Numerical tests for GPU acceleration without dynamic parallelism}
\begin{figure}
    \centering
    \includegraphics[scale=0.3]{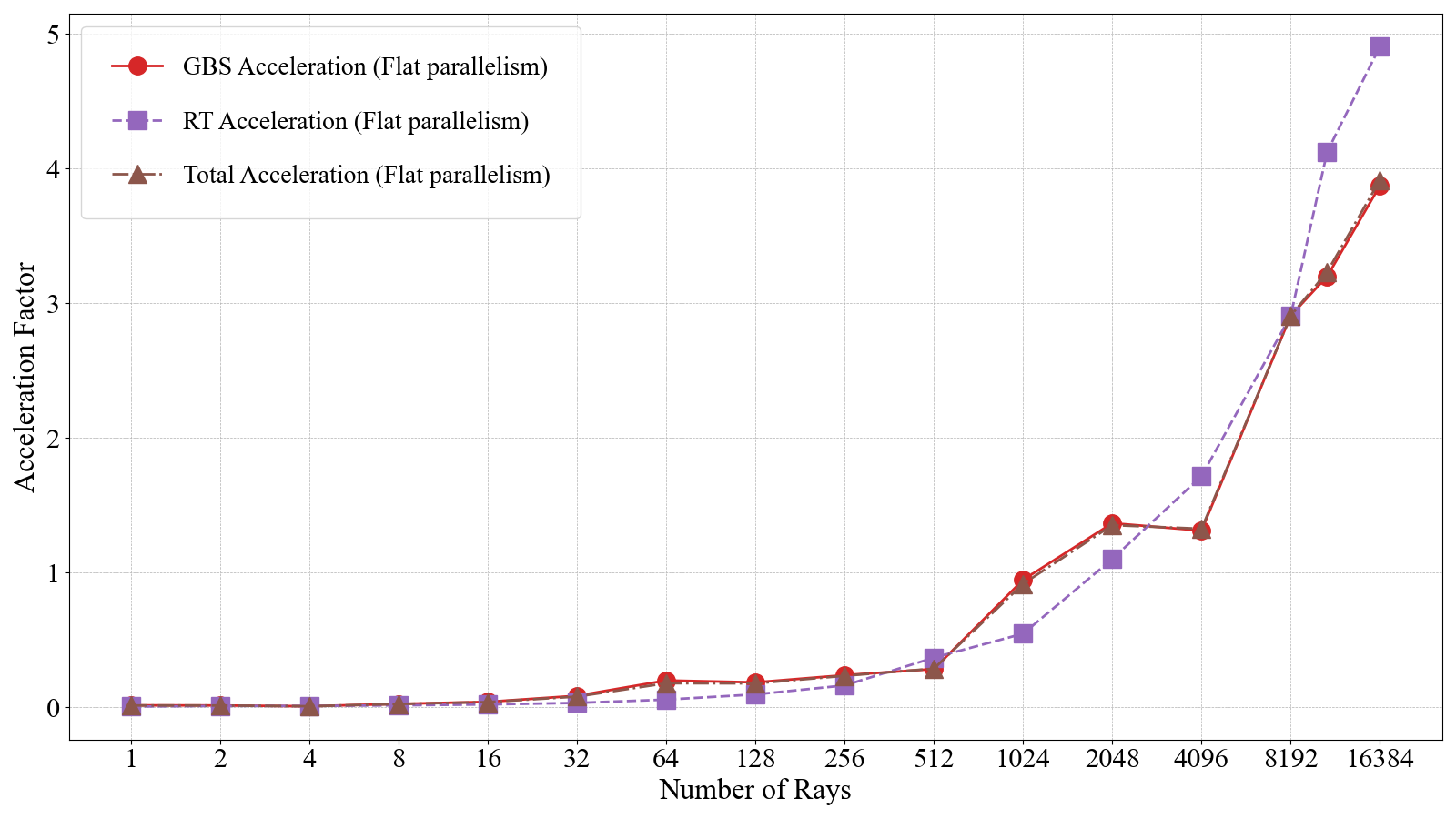}
    \caption{Performance comparison of multithreaded sound field prediction using flat parallelism GPUs: Acceleration factor for different ray counts.}
    \label{fig:Performance of GPU multithreaded}
\end{figure}

\begin{figure}
    \centering
    \includegraphics[scale=0.30]{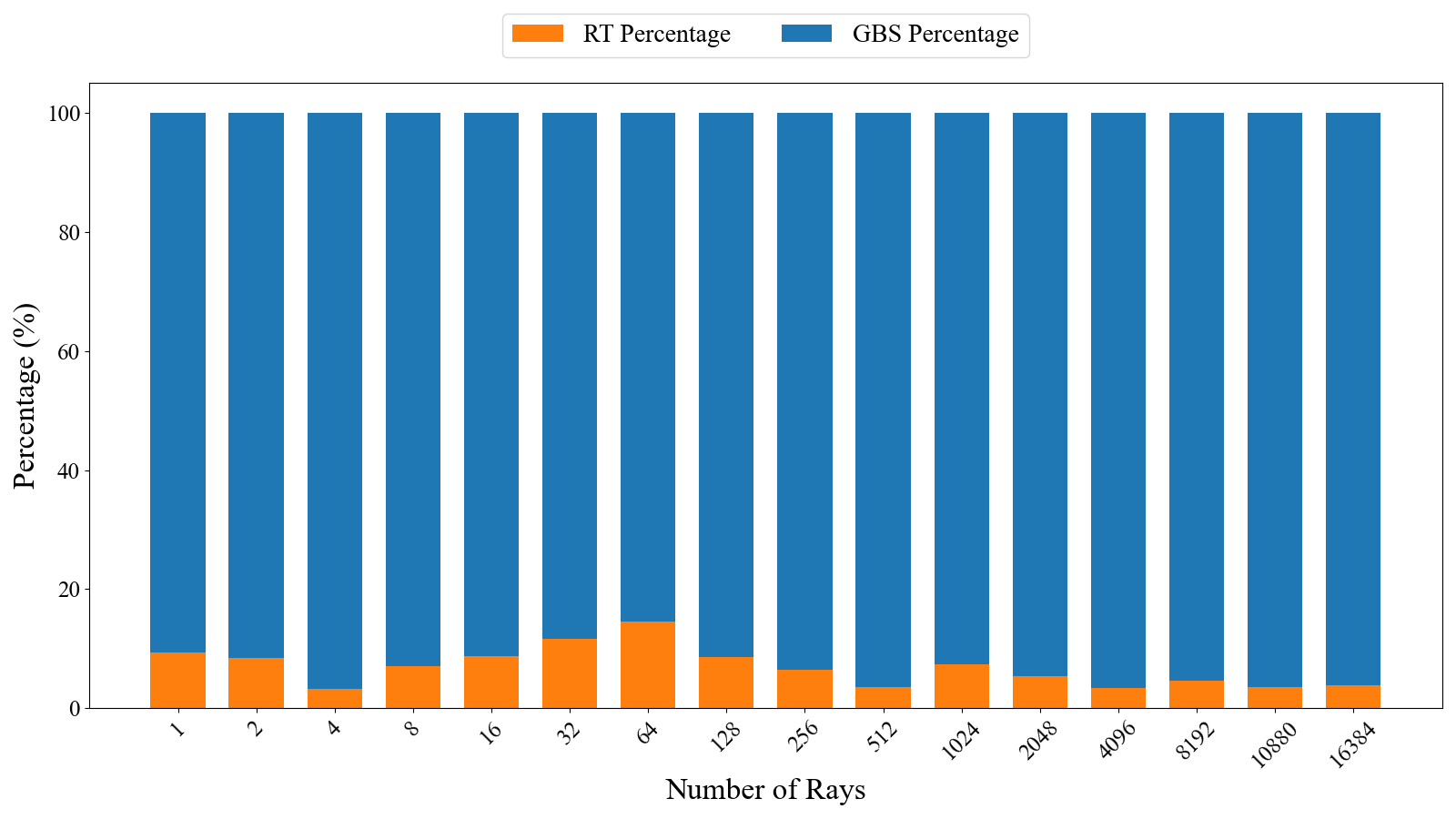}
    \caption{Percentage of two main processes in a multi-threaded sound field prediction algorithm using flat parallelism GPUs.}
    \label{fig:Percentage of two major processes}
\end{figure}

In the preliminary phase of this study, the focus is on migrating existing single-threaded sound field prediction algorithms to the GPU platform, aiming to enhance computational efficiency through hardware acceleration. The experimental scenario involves complex city environments. Experimental results demonstrate that, compared to sequential execution on a CPU, the acceleration effect achieved by leveraging GPU capabilities is illustrated in fig.~\ref{fig:Performance of GPU multithreaded} and fig.~\ref{fig:Percentage of two major processes}. These figures likely showcase significant improvements in processing time and efficiency, highlighting the benefits of utilizing GPU acceleration for sound field prediction in complex enviroment.

In the experiments, we observed that as the number of rays increased, the acceleration factor for GBS gradually improved, starting from an initial factor of 0.013 and increasing up to 3.47. This indicates that GPU acceleration becomes increasingly efficient in processing GBS computations as the number of rays grows. Similarly, the acceleration factor for Ray Tracing (RT) also showed a certain growth trend, increasing from an initial factor of 0.04 up to 4.90.

With the increase in the number of rays, both the GBS and RT acceleration factors exhibited improvements. Given that the GBS part occupies more than 95\% of the total computation time, the overall acceleration factor is close to that of GBS. These results indicate that despite the observed acceleration effect, the overall acceleration performance is still limited due to the influence of the number of rays on the number of parallel GPU threads and the fact that the single-thread performance of GPUs is much lower compared to CPUs.

To further analyze the GPU performance utilization, we employ NVIDIA's performance analysis tool, Nsight Compute. This tool can provide detailed GPU performance metrics, helping us understand the efficiency of the program's execution on the GPU. The analysis results are presented in Table 1.
\begin{table}[h]
\centering
\caption{GPU Speed Of Light Throughput Metrics}
\begin{tabular}{lc}
\hline
\textbf{Metric} & \textbf{Throughput [\%]} \\
\hline
Compute (SM) Throughput & 3.23 \\

Memory Throughput & 1.68 \\

L1/TEX Cache Throughput & 24.64 \\

L2 Cache Throughput & 1.68 \\

DRAM Throughput & 0.50 \\
\hline
\end{tabular}
\label{tab:gpu_speed_of_light_throughput}
\end{table}

The Compute (SM) Throughput of 3.23\% represents the efficiency of the Streaming Multiprocessors (SMs) on the GPU when executing computational tasks relative to their theoretical maximum efficiency. This low percentage indicates that the utilization of computational resources is not high, and improving their utilization is a top priority.

Although the use of the GPU platform has achieved computational acceleration to some extent, the efficiency gains are limited when dealing with complex acoustic environments and high-density ray grids. Especially in enviroment involving high-frequency sound waves and complex environmental interactions, relying solely on the hardware advantages of GPUs has not fully addressed the performance bottlenecks in sound field prediction, leading to significant limitations in computational efficiency.

\subsection{Numerical test for GPU acceleration with dynamic parallelization}
After implementing dynamic parallel optimization for the sound field prediction algorithm, we conducted a detailed evaluation of the algorithm's acceleration performance, with a particular focus on the performance improvement in the GBS section after introducing dynamic parallel processing. To ensure the effectiveness of the proposed algorithm and the accuracy of the performance evaluation, the hardware configuration included an NVIDIA RTX A6000 GPU and the CUDA version used was 12.03. This setup aimed to provide an efficient, stable, and reproducible experimental platform for evaluating the proposed multi-threaded sound field prediction algorithm. With such a configuration, the experimental results can effectively reflect the algorithm's performance and provide valuable references for subsequent research and applications.

To evaluate the performance of the proposed multi-threaded sound field prediction algorithm, the program was executed under two different complexity enviroment - “Free Field” and “city enviroment”. In the experiments, we compare the performance of executing the same tasks using a single-threaded CPU and a single-piece GPU, paying special attention to the time proportion of the GBS and RT sections when processing different numbers of rays, as well as the acceleration factor of the GPU relative to the CPU. The GPU utilization is also analyzed using NVIDIA's Nsight Compute tool.
\subsection{Discussion of two major examples}

\begin{table}[h]
\centering
\caption{Free field case information with detailed descriptions}
\begin{tabular}{lc}
\hline
\textbf{Parameter} & \textbf{Value} \\
\hline
$T_a$: Air temperature (\textdegree C) & 20 \\
$H_r$: Relative humidity (\%) & 70 \\
$P_a$: Atmospheric pressure (atm) & 1 \\

$f_s$: Number of source frequencies & 5 \\
$\text{Im}(b)$: Imaginary component of beam parameter & -45874 \\

$N_b$: Total number of buildings & 1000 \\
$N_t$: Total number of terrain grid points & 4000 \\
$N_r$: Total number of roads & 5000 \\
$N_w$: Number of water bodies (negative for exclusion) & -10 \\
$N_{\text{tree}}$: Number of trees (negative for exclusion) & -10 \\

$D$: Simulation dimensionality (3D or 2D) & 3D \\
$\Theta_{\min}$: Minimum elevation angle & 0 \\
$\Theta_{\max}$: Maximum elevation angle & 180 \\
$\Phi_{\min}$: Minimum azimuthal angle & 0 \\
$\Phi_{\max}$: Maximum azimuthal angle & 360 \\

$N_{\text{steps}}$: Total time steps & 8000 \\
$R_{\max}$: Maximum number of sound reflections allowed & 10 \\
$\Delta t$: Duration of a single time step (s) & 0.0001 \\
$N_o$: Total number of observation points & 13586 \\
\hline
\end{tabular}
\label{tab:free_field_case}
\end{table}

In the simulation of this algorithm, atmospheric conditions including air temperature, relative humidity, and atmospheric pressure are assumed. These parameters are key factors affecting sound wave propagation, as they directly influence the speed of sound and its interaction with the environment. The imaginary part of the single sound source frequency and beam parameters in the GBT data is an important component of the analysis, helping us construct a more realistic sound propagation model. Moreover, by setting different observation angles ($\Theta$ and $\Phi$ representing elevation and azimuth angles, respectively) and time steps (e.g., 5000 time steps, each step 0.0001 seconds), we are able to capture the propagation paths of sound waves at a higher resolution. The flow impedance setting for environmental elements is set to -10, indicating that all walls and obstacles are considered as hard surfaces, meaning that sound waves will reflect off them instead of penetrating. This setting is based on the characteristics of building materials commonly found in city environments. The observation points set in the model not only show the spatial distribution of sound wave propagation but also provide a comprehensive perspective for sound field analysis. The specific input information is illustrated in Tables 2 and 3.

\begin{table}[h]
\centering
\caption{City case information with detailed descriptions}
\begin{tabular}{lc}
\hline
\textbf{Parameter} & \textbf{Value} \\
\hline
$T_a$: Air temperature (\textdegree C) & 20 \\
$H_r$: Relative humidity (\%) & 70 \\
$P_a$: Atmospheric pressure (atm) & 1 \\

$f_s$: Number of source frequencies & 1 \\
$\text{Im}(b)$: Imaginary component of beam parameter & -45874 \\

$N_b$: Total number of buildings & -10 \\
$N_t$: Total number of terrain grid points & -10 \\
$N_r$: Total number of roads & -10 \\
$N_w$: Number of water bodies (negative for exclusion) & -10 \\
$N_{\text{tree}}$: Number of trees (negative for exclusion) & -10 \\

$D$: Simulation dimensionality (3D or 2D) & 3D \\
$\Theta_{\min}$: Minimum elevation angle & 0 \\
$\Theta_{\max}$: Maximum elevation angle & 180 \\
$\Phi_{\min}$: Minimum azimuthal angle & 0 \\
$\Phi_{\max}$: Maximum azimuthal angle & 360 \\

$N_{\text{steps}}$: Total time steps & 5000 \\
$R_{\max}$: Maximum number of sound reflections allowed & 20 \\
$\Delta t$: Duration of a single time step (s) & 0.0001 \\
$N_o$: Total number of observation points & 10201 \\
\hline
\end{tabular}
\label{tab:city_case}
\end{table}

In the "City environment" case, acoustic simulations are conducted for a complex city environment. As illustrated in Fig.~4.1, the model integrates a variety of environmental elements, including buildings, terrain, and roads. The combination of these elements not only reflects the diversity of city environments but also provides realistic simulation conditions for the propagation of sound waves through multiple obstacles. In contrast, the “Free Field” case study involves simulations in an open space environment without obstacles, hence it does not involve the setting of complex environmental elements. This distinction between the two enviroment allows for a comprehensive analysis of sound propagation behaviors in different settings, highlighting the effects of city structures on acoustic phenomena.

\subsubsection{Numerical analysis of city enviroment}
\begin{figure}
    \centering
    \includegraphics[scale=0.3]{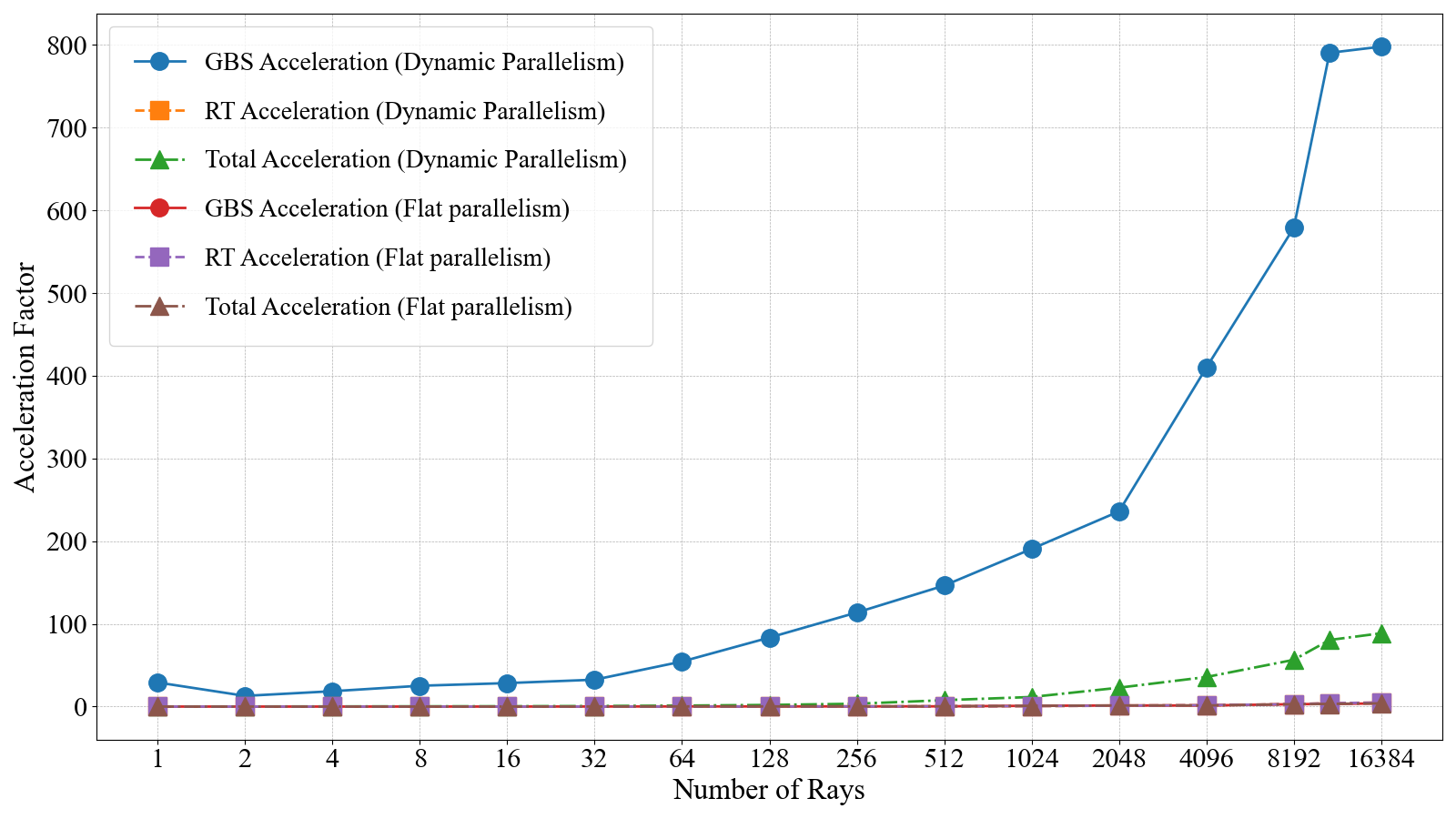}
    \caption{Performance of multithreaded GPU sound field prediction in city environments using dynamic parallelism: Acceleration factors for different ray counts.}
    \label{fig:Performance of optimised GPU multithreaded}
\end{figure}

\begin{figure}
    \centering
    \includegraphics[scale=0.30]{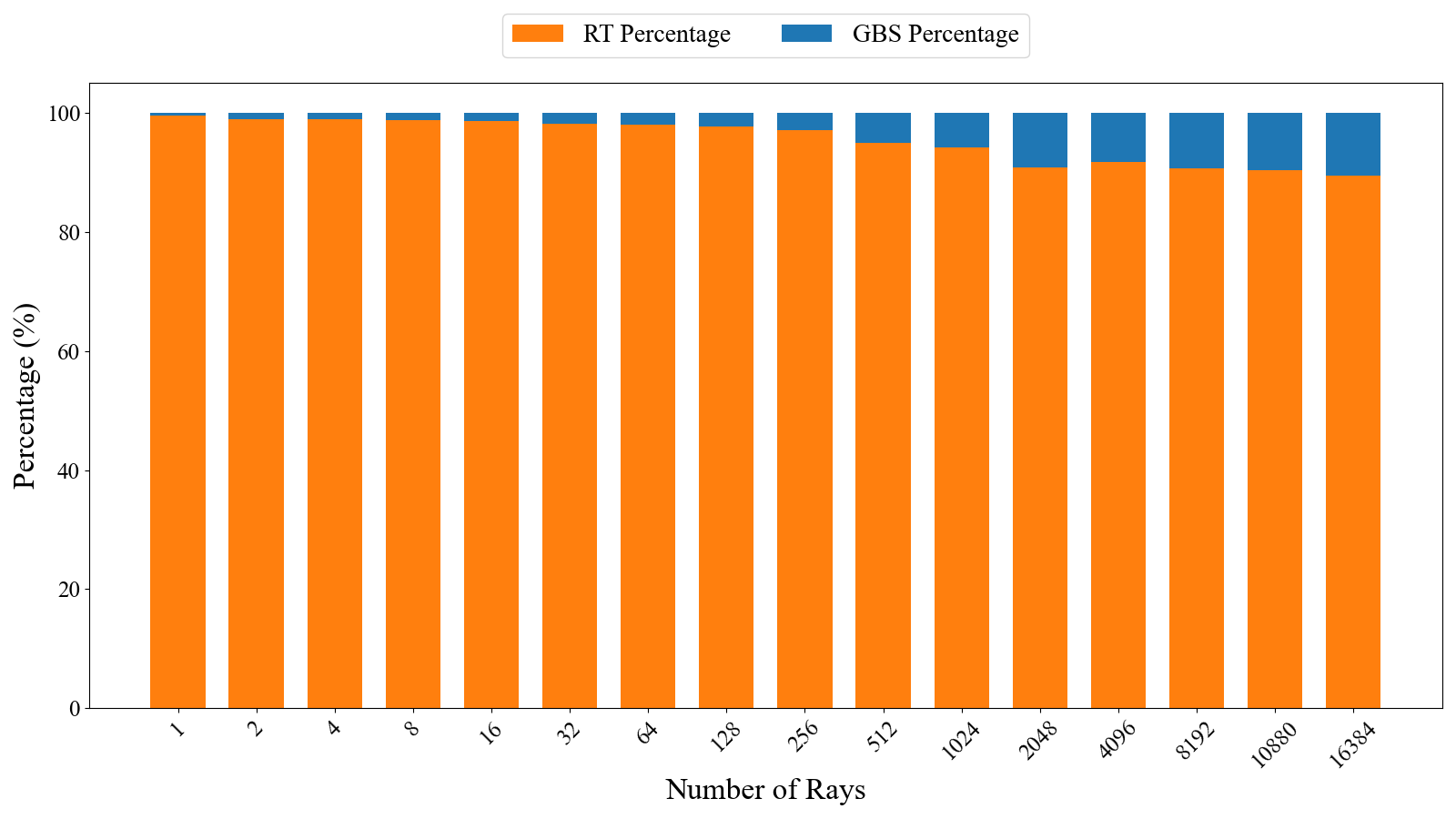}
    \caption{Percentage of two main processes in a multi-threaded sound field prediction algorithm in city environments using dynamic parallelism GPUs.}
    \label{fig:Percentage of the two main processes in the optimized GPU multithreaded sound}
\end{figure}

Our GPU multithreaded algorithm demonstrated significant acceleration compared to the traditional CPU single-threaded algorithm in the “city enviroment” experiment. This is particularly evident when the number of rays is increased to 16384, with the acceleration of the GBT reaching an impressive 798 times and the overall acceleration reaching approximately 88 times. These remarkable results are clearly illustrated in Figure.~\ref{fig:Performance of optimised GPU multithreaded}. The increased thread-parallel computing power provided by GPU dynamic parallelism is the main reason for the significant performance improvement. Furthermore, the timeshare analysis of the GBS and RT components clearly identifies a performance bottleneck in the sound field prediction algorithm. In multi-threaded GPU execution, the time taken by the GBS component is significantly reduced to the extent that RT occupies more than 90\% of the time, as shown in Figure.~\ref{fig:Percentage of the two main processes in the optimized GPU multithreaded sound}. This highlights the significant advantage of GPU dynamic parallelism in processing complex acoustic computations. The overall acceleration ratio is nearly an order of magnitude greater than the acceleration ratio of the GBS component. The limitation of parallel acceleration potential is solely due to the serial execution of the RT part of the computation in a single thread. Nevertheless, the GPU multithreaded sound field prediction algorithm demonstrates significant acceleration, up to a hundred times in complex enviroments.

During the “City enviroment” experiment, it is observed that the GPU's memory capacity limited the number of rays that could be processed simultaneously. The maximum ray processing capacity is determined to be 11,364 rays, while the experiment involved 16,384 rays, exceeding this limit. To address this issue, a chunking mechanism is introduced, which involves dividing the large-scale ray computation task into multiple smaller chunks to accommodate the memory capacity constraints. This mechanism allows the GPU to effectively handle several rays that exceed its one-time maximum processing capability. Notably, after implementing the chunk processing operation, the acceleration factor for GBS reached 798 times, which is close to the 790 times acceleration factor observed with 10,800 rays previously. This result indicates that, despite the introduction of chunk processing adding additional overhead for memory management and computation scheduling, the acceleration factor remains at a high level and tends to converge with the increase in the number of processing chunks. This phenomenon reveals the stability and efficiency of the GPU multi-threaded algorithm in handling large-scale sound field prediction tasks.

\subsubsection{Numerical analysis of free-field enviroment}
\begin{figure}
    \centering
    \includegraphics[scale=0.3]{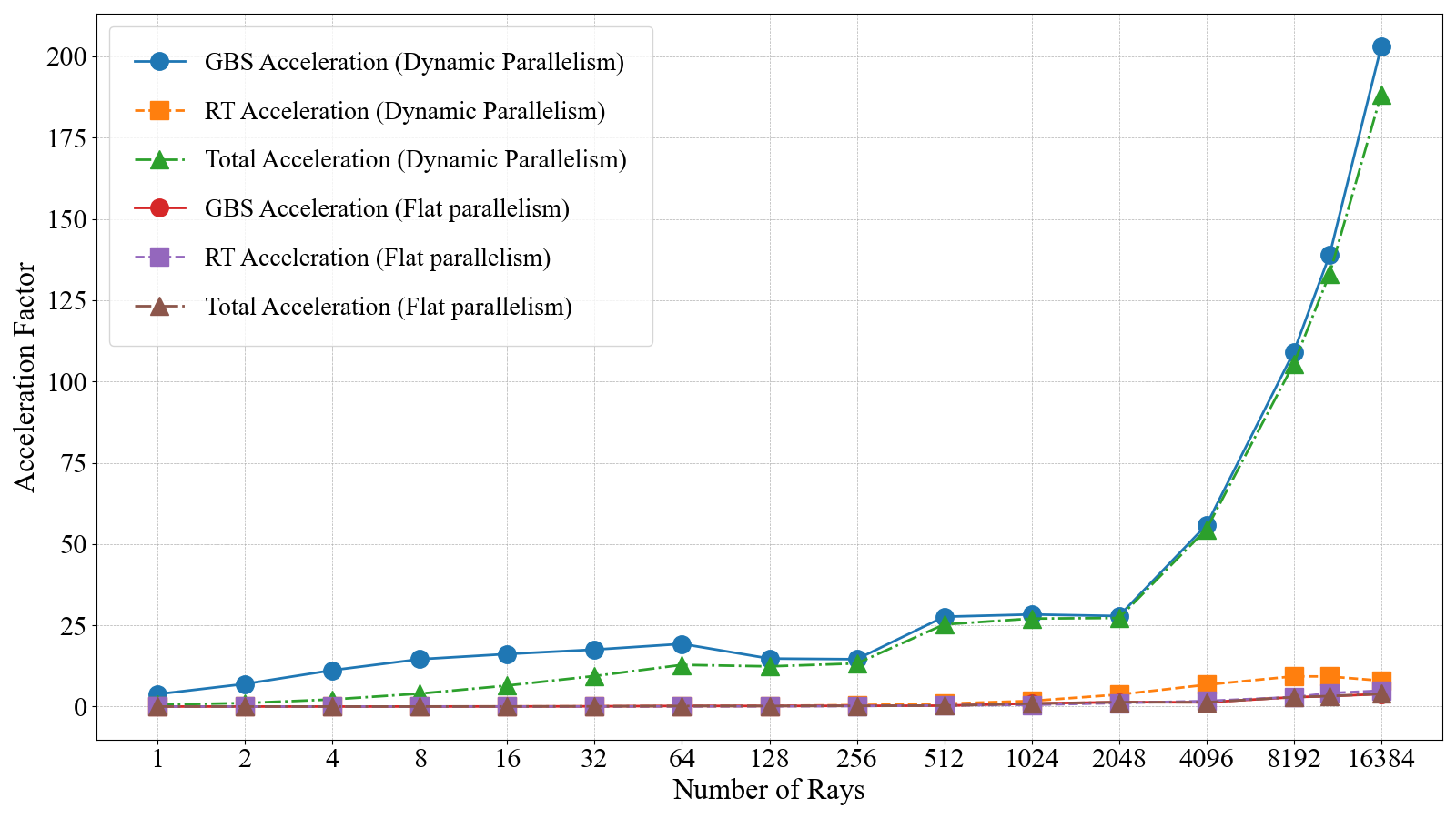}
    \caption{Performance of multithreaded GPU sound field prediction in free field environments using dynamic parallelism: Acceleration factors for different ray counts.}
    \label{fig:Performance of optimised GPU multithreaded sound field}
\end{figure}
\begin{figure}
    \centering
    \includegraphics[scale=0.30]{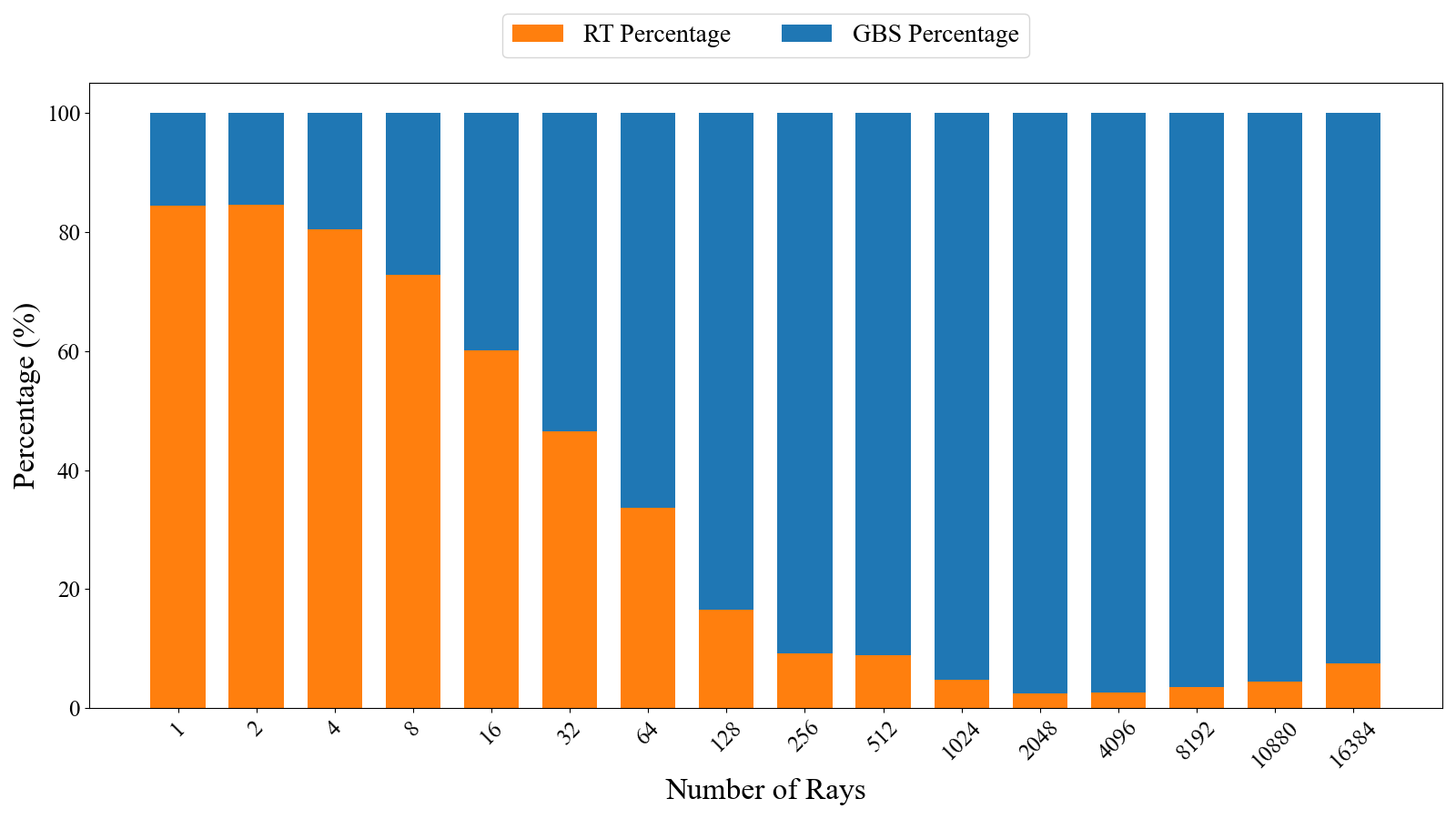}
    \caption{Percentage of two main processes in a multi-threaded sound field prediction algorithm in free field environments using dynamic parallelism GPUs.}
    \label{fig:Percentage of the two main processes in the optimized GPU multithreaded}
\end{figure}

The "Free Field" scenario exhibit particularly noteworthy performance when the number of rays increased to 16,384. Although the acceleration factor for GBS  only reached 200 times, the overall acceleration factor significantly rose to 188 times, as illustrated in Fig.~\ref{fig:Performance of optimised GPU multithreaded sound field}. A shift in the time proportion between the RT module and the GBS module occurred, as shown in Fig.~\ref{fig:Percentage of the two main processes in the optimized GPU multithreaded}, where the roles reversed in terms of their contribution to the total computation time.

The unusual bar chart highlights the performance dynamics of the sound field prediction algorithm across different enviroment. Initially, the RT phase occupies a higher proportion of the computational load, while the GBS phase remains relatively low. This reason is that, with a smaller number of rays, the GBS phase has fewer Gaussian beams to sum, resulting in lower computational demand. However, as the number of rays increases, the workload for GBS escalates significantly due to the larger number of beams that need to be processed. In contrast, the RT phase experiences less pressure increase, especially in a free field scenario where there are no complex architectural elements to trace. This results in the chart showing a higher proportion of RT initially, followed by a dominant GBS proportion as the ray count grows. This pattern effectively illustrates the shifting computational burden between RT and GBS as the complexity of the ray interactions increases.

\subsubsection{Analysis of GPU Utilisation Resource Levels}
The analysis of GPU resource utilization is conducted using Nsight compute during the city scenario experiment. The results are presented in Tables 4, 5, and 6.
\begin{table}[h]
\centering
\caption{Optimized Nsight compute software performance analysis table}
\begin{tabular}{lc}
\hline
\textbf{Metric} & \textbf{Throughput [\%]} \\
\hline
Compute (SM) Throughput & 22.27 \\

Memory Throughput & 9.16 \\

L1/TEX Cache Throughput & 13.15 \\

L2 Cache Throughput & 9.16 \\

DRAM Throughput & 2.04 \\
\hline
\end{tabular}
\label{tab:gpu_speed_of_light_throughput}
\end{table}

A significant change observed in Table 4 is the substantial increase in the utilization of Streaming Multiprocessors (SM) resources from a mere 3.23\% to 22.27\% by implementing dynamic parallel processing for the GBS section. This significant improvement indicates that the dynamic parallelism strategy plays a key role in enhancing the efficiency of GPU computational resource utilization.
\begin{table}[h]
\centering
\caption{Performance analysis of the Nsight compute software for Ray tracing process graph}
\begin{tabular}{lc}
\hline
\textbf{Metric} & \textbf{Throughput [\%]} \\
\hline
Compute (SM) Throughput & 1.67 \\

Memory Throughput & 3.05 \\

L1/TEX Cache Throughput & 4.38 \\

L2 Cache Throughput & 3.05 \\

DRAM Throughput & 0.32 \\
\hline
\end{tabular}
\label{tab:gpu_speed_of_light_throughput}
\end{table}

As seen in Table 5, the complexity of the computation leads to a higher demand for registers per GPU thread, with an average of 199 registers needed. This high register demand poses a limitation on the granularity of dynamic parallelism. Registers are a highly valuable resource in GPUs, and the excessive consumption of registers by each thread means that fewer threads can be executed simultaneously. Therefore, our dynamic parallelism optimization strategy is approaching its performance limit under the current conditions.

Table 6 shows that the GPU utilization rate for executing the RT module alone is only 1.67\%, a low utilization rate that significantly drags down the overall GPU performance. This indicates significant room for performance improvement during the RT process. Thus, a focus of future work could be to further optimize the RT algorithm to improve its resource utilization efficiency on the GPU, thereby achieving a higher acceleration ratio and better performance for the overall sound field prediction algorithm.

\begin{table}[h]
\centering
\caption{Analysis of the resources used by the optimised}
\begin{tabular}{lc}
\hline
\textbf{Launch Statistics} & \textbf{Value} \\
\hline
Grid Size & 6 \\

Registers Per Thread & 199 \\

Block Size & 256 \\

Thread & 1536 \\

Waves Per SM & 0.13 \\
\hline
\end{tabular}
\label{tab:gpu_speed_of_light_throughput}
\end{table}

\section{Conclusion}

This study successfully reconstructed the Gaussian Beam Tracing (GBT) method using CUDA, enabling efficient GPU acceleration for complex acoustic simulations. By leveraging both flat and dynamic parallelism, the proposed method addresses computational challenges such as irregular loops and GPU memory limitations, significantly improving performance and scalability. The key findings and contributions of this work are as follows:

Substantial performance gains: The GPU-accelerated GBT algorithm demonstrated significant acceleration, achieving up to 790 times speedup in city environments and 188 times speedup in free-field environments compared to the original CPU-based implementation. These results establish the effectiveness of the proposed method in handling large-scale acoustic simulations in diverse scenarios.

Innovative algorithmic enhancements: This study introduced a dynamic parallelism approach to efficiently process the non-uniform computational load of Gaussian beam summation (GBS). By enabling threads to launch new threads dynamically, the method effectively balanced workloads, minimized idle resources, and overcame the limitations of flat parallelism in scenarios with highly variable computational demands.

Comprehensive evaluation: Extensive numerical experiments validated the accuracy of the GBT method against analytical solutions and assessed its performance in both free-field and city environments. The proposed chunking mechanism for managing GPU memory further ensured scalability for processing a large number of rays.

Scalability and practical applications: The findings highlight the potential for real-time sound field prediction in applications such as architectural acoustics, environmental noise analysis, and virtual reality. The integration of GPU-accelerated algorithms offers a robust foundation for future developments in high-performance acoustic simulations.

\section*{Acknowledgments}
This work was supported in part by the National Natural Science Foundation of China (No. 12302346) and
Zhejiang Provincial Natural Science Foundation of China (No. LQ24A040014).

\bibliographystyle{unsrt}
\bibliography{Refs}
\end{document}